\documentclass[a4paper,11pt]{article}
\usepackage{pos}

\title{Dark matter searches at LHCb}

\author*[a,\dagger]{Titus Momb\"acher}

\affiliation[a]{Instituto Galego de Física de Altas Enerxías,\\
  Rúa de Xoaquín Díaz de Rábago, s/n, Santiago de Compostela, Spain}

\note[$\dagger$]{On behalf of the LHCb collaboration}

\emailAdd{titus.mombacher@cern.ch}

\abstract{In the extensive efforts to understand the nature of Dark Matter and searches at colliders, the LHCb experiment has a unique sensitivity to low mass Dark Matter candidates. These proceedings present recent results and prospects on Dark Matter searches with the LHCb experiment that achieve world-leading sensitivities.}

\FullConference{%
  *** The European Physical Society Conference on High Energy Physics (EPS-HEP2021), ***\\
  *** 26-30 July 2021 ***\\
  *** Online conference, jointly organized by Universität Hamburg and the research center DESY ***
}

\begin{document}
\maketitle

The origin and nature of Dark Matter (DM) are among the most pressing unknowns in the field of particle physics.
Many different models proposed to describe it introduce signatures that can be searched for at colliders.
These typically arise through so-called portals to possible Dark Sectors (DS), new mediators that mix with the Standard Model (SM) and couple to DM candidates.
Dark Sector portals allow either direct DM production from SM particles, which would lead to missing energy signatures at colliders, or the production and decay of new particles into SM particles. With a weak mixing between the new mediators and SM mediators, the latter decays can be long lived and thus avoid many experimental bounds as most searches target prompt decays.

The LHCb experiment with its acceptance region at high pseudorapidity and excellent vertex and mass resolution provides a unique coverage of the aforementioned signatures that is complementary to the other collider experiments such as ATLAS and CMS. In particular it allows the access of signatures with low masses and significant displacement.

These proceedings review a subset of recent LHCb Dark Matter searches and gives prospects for the upcoming data-taking periods.
\section{Search for Dark Photons through \boldmath$A^\prime\to\mu^+\mu^-$}
A popular way to connect the SM to a possible DS is through massive Dark Photons $A^\prime$ that can kinetically mix with off-shell photons $\gamma^*$~\cite{DM2016}.
The LHCb experiment performed a search for these Dark Photons through the decay $A^\prime\to\mu^+\mu^-$ using prompt and displaced signatures with $pp$-collision data collected at a centre-of-mass energy of $13\,\text{TeV}$ in 2016-2018 corresponding to an integrated luminosity of about $5.5\,\text{fb}^{-1}$~\cite{DarkPhoton1, DarkPhoton2}. The search utilises the similar $pp\to\mu^+\mu^-$ production through the off-shell photon as normalisation mode, which allows the exact cancellation of almost all systematic uncertainties via 
\begin{align}
n^{A^\prime}_{\text{ex}}[m(A^\prime),\varepsilon^2]=\varepsilon^2{\left[\frac{n^{A^\prime}_{\text{ob}}[m(A^\prime)]}{2\Delta m}\right]}{\mathcal{F}[m(A^\prime)]}{\epsilon^{A^\prime}_{\gamma^*}[m(A^\prime),\tau(A^\prime)]}
\end{align}
(modifications according to mixing with the $Z$ apply at high dimuon masses).
Here, $n^{A^\prime}_{\text{ex}}$ are the expected number of Dark Photon decays in a dimuon mass window $\pm\Delta m$ and $n^{A^\prime}_{\text{ob}}$ the observed number of off-shell photon decays in that window.
The only differences between the two decay modes arise from the kinetic mixing strength between the Dark Photon and the off-shell SM photon $\varepsilon^2$, a phase-space factor $\mathcal{F}$ from the mass of the hypothetical Dark Photon $m(A^\prime)$ and the efficiency for the displaced searches that depends on the Dark Photon mass and lifetime $\tau(A^\prime)$.
This allows a fully data-driven analysis for the prompt decays based on online-selected data~\cite{Trigger_Performance}. The search for displaced decays, however, uses offline selected data.
The search is performed as a bump hunt over a large dimuon mass ranging from $m(\mu^+\mu^-)\in[0.7,70]\,\text{GeV}/c^2$ for prompt decays and $m(\mu^+\mu^-)\in[214,350]\,\text{MeV}/c^2$ for displaced decays that correspond to a lifetime of $\mathcal{O}(1\,\text{ps})$.
A significant challenge in this analysis is the suppression of background and precise estimation of its yield.
Backgrounds for the prompt search arise from prompt $\gamma^*\to\mu^+\mu^-$ decays, which are irreducible, midentified hadrons, as well as muons from heavy flavour decays. The latter two categories are heavily suppressed through muon identification criteria and prompt-like requirements. Furthermore, dimuon regions of known resonances are vetoed. For $m(\mu^+\mu^-)\gtrsim 1\,\text{GeV}/c^2$ the signal process is expected to be Drell-Yan-like and therefore the two muons isolated from the remainder of the event. Thus, above the $\phi$-resonance, anti-$k_\text{T}$-based isolation criteria are applied. The composition of the backgrounds is measured in data with an extended binned maximum likelihood fit to the dimuon vertex fit quality and the minimum $\chi^2_{\text{IP}}$ of the two muons, where the latter is defined as the difference in the vertex fit quality of the primary vertex with and without the muon candidate.

For the long-lived search, dominant backgrounds stem from photon conversions through material interactions inside the Vertex Locator detector of the experiment, heavy flavour decays and doubly misidentified $K^0_\text{S}\to\pi^+\pi^-$ decays, which each can be strongly reduced through a p-value based on the material map of the detector, efficient track isolation criteria and stringent muon identification requirements.

\begin{figure}
\centering
\includegraphics[width=0.7\textwidth]{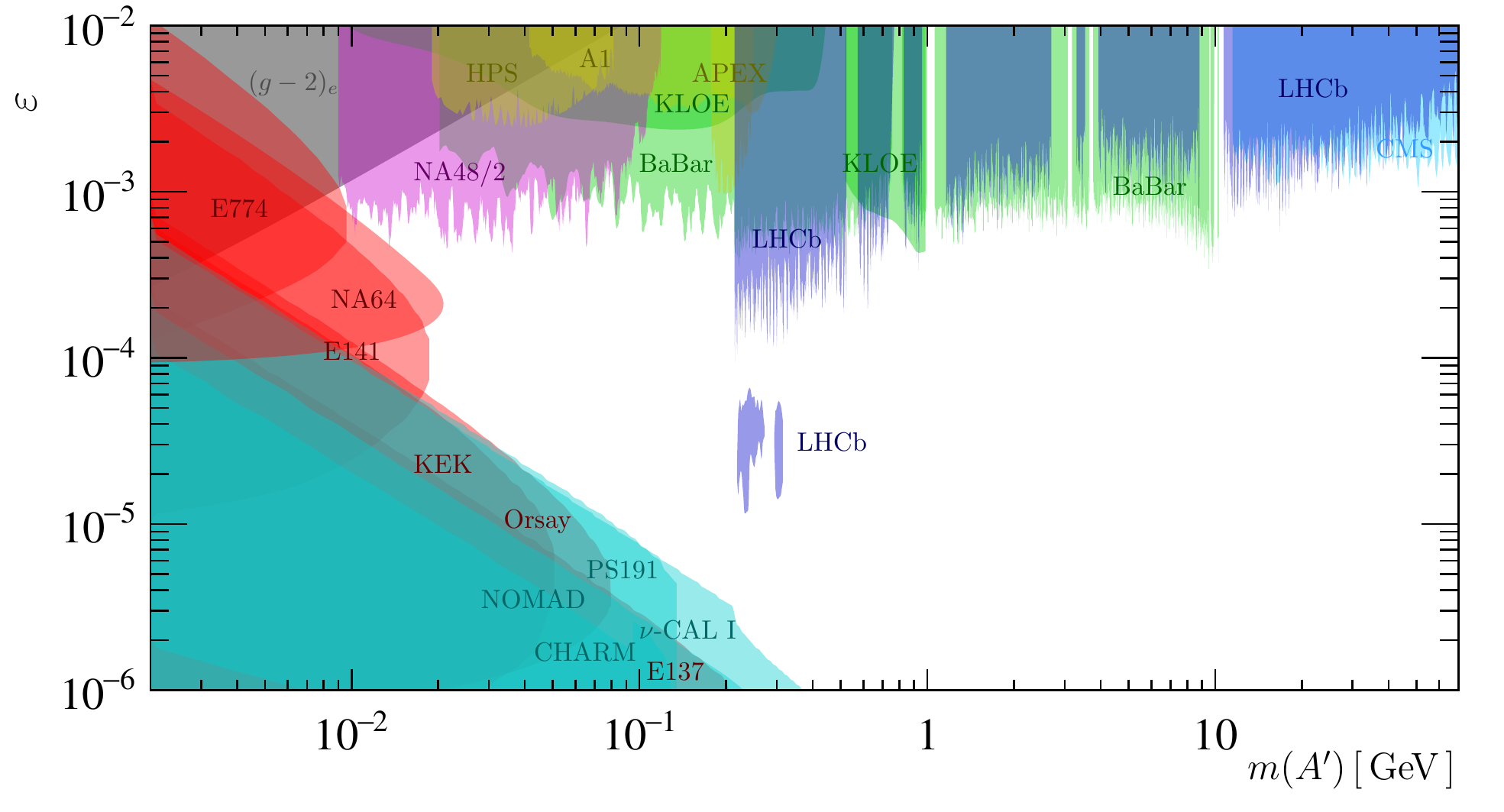}
\caption{Exclusion regions in the plane of the kinetic mixing strength $\varepsilon^2$ and the Dark Photon mass $m(A^\prime)$ including the recent LHCb result~\cite{DarkPhoton2}. The LHCb search for displaced dimuons covers a region far beyond the sensitivity of any other experiment.}
\label{fig:dark_photon}
\end{figure}

From the bump hunt limits on the strength of the kinetic mixing with the off-shell photon are obtained and shown in Fig.~\ref{fig:dark_photon}. They are the most stringent ones for $214<m(A^\prime)\lesssim740\,\text{MeV}/c^2$ and $10.6<m(A^\prime)\lesssim30\,\text{GeV}/c^2$. The long-lived search covers a fully unexplored region at very low mixing strengths.
With the upcoming data-taking period starting in 2022, even lower masses for the Dark Photon will be explored through $D^{*0}\to D^0A^\prime(e^+e^-)$ decays with the possibility of excluding mixing strengths down to $\varepsilon^2\lesssim10^{-10}$~\cite{DarkPhoton_e}.

\section{Inclusive search for \boldmath$X\to\mu^+\mu^-$}
\begin{figure}
\centering
\includegraphics[width=0.7\textwidth]{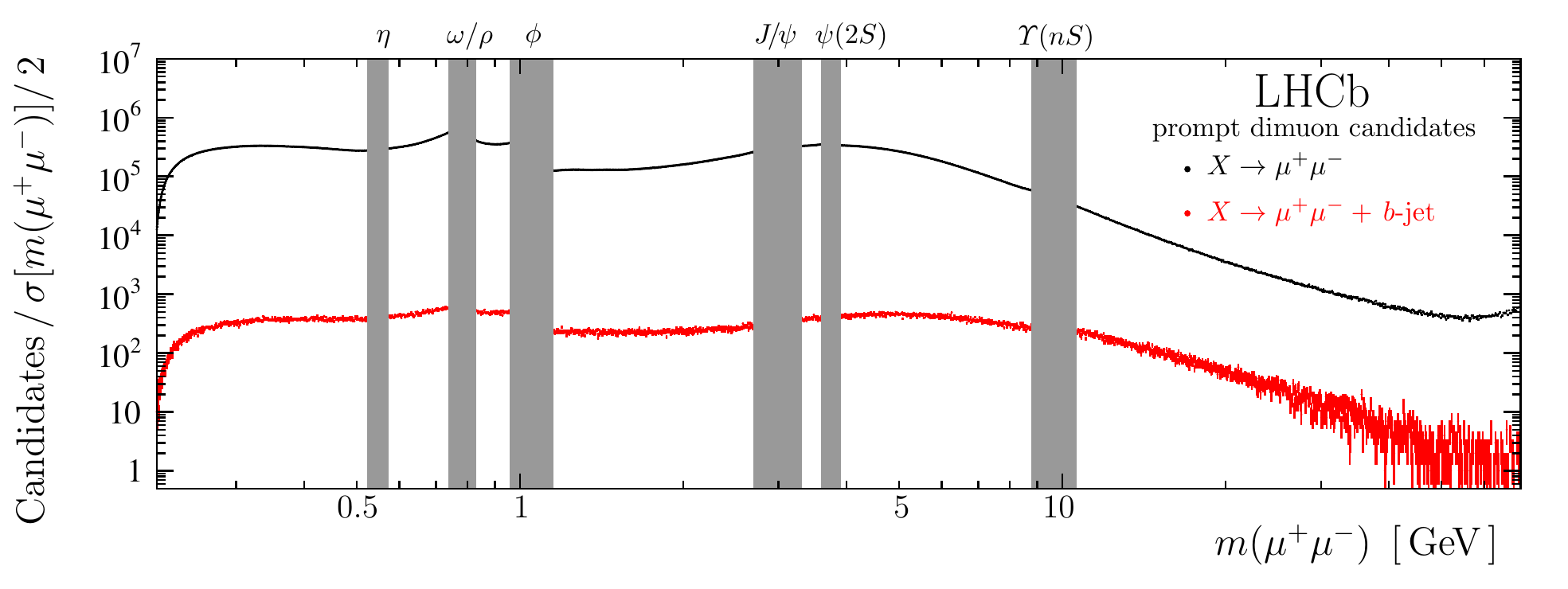}
\includegraphics[width=0.7\textwidth]{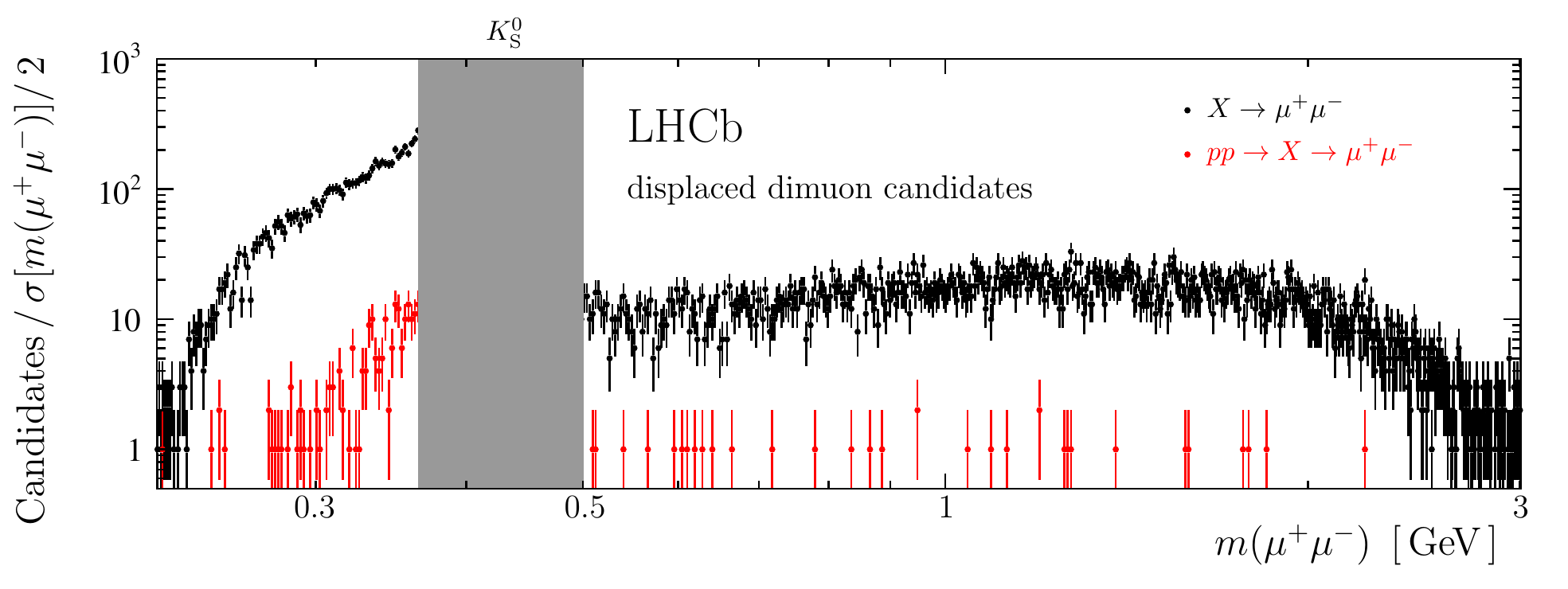}
\caption{Dimuon invariant-mass spectra of the inclusive search for $X\to\mu^+\mu^-$ with the LHCb experiment~\cite{Xmumu}. The grey areas display the regions of known resonances and are vetoed.}
\label{fig:dimuon_spectra}
\end{figure}

Since there is no clearly favoured DM model, an attempt is made to generalise the search for dimuons into a model-independent analysis with a slightly reduced data set corresponding to an integrated luminosity of $5.1\,\text{fb}^{-1}$~\cite{Xmumu}. Four different decay topologies are considered for a hypothetical particle decaying into at least two muons: inclusive prompt dimuon decays; inclusive prompt dimuon decays that are associated with a $b$-jet as defined with an anti-$k_\text{T}$ algorithm with $R=0.5$ and where the jet cone contains a secondary vertex; displaced dimuon decays pointing and non-pointing to the primary vertex.
Similarly to the Dark Photon search, a bump hunt is performed over a large dimuon invariant mass, where known resonances are vetoed. For prompt decays above $m(\mu^+\mu^-)>20\,\text{GeV}/c^2$ non-negligible widths of up to $\Gamma(X)\leq3\text{GeV}$ are considered. The dimuon spectra are shown in Fig.~\ref{fig:dimuon_spectra}.

The search is performed in tight fiducial regions to facilitate the interpretation in any model that would produce a promptly decaying low-mass dimuon resonance within these regions. No excess is found and upper limits on the production cross-sections are set.
If interpreted \textit{e.g.} in a model with a two-Higgs doublet potential where a scalar singlet is added~\cite{2HDM}, the results on the prompt-like searches impose most stringent bounds on the model, while ruling out an excess seen by CMS in a dimuon search associated with a $b$-jet~\cite{CMS_mmb} with a $20$ times stronger limit.

\section{Prospects on searches for baryonic Dark Matter}
\begin{figure}
\centering
\includegraphics[width=0.49\textwidth]{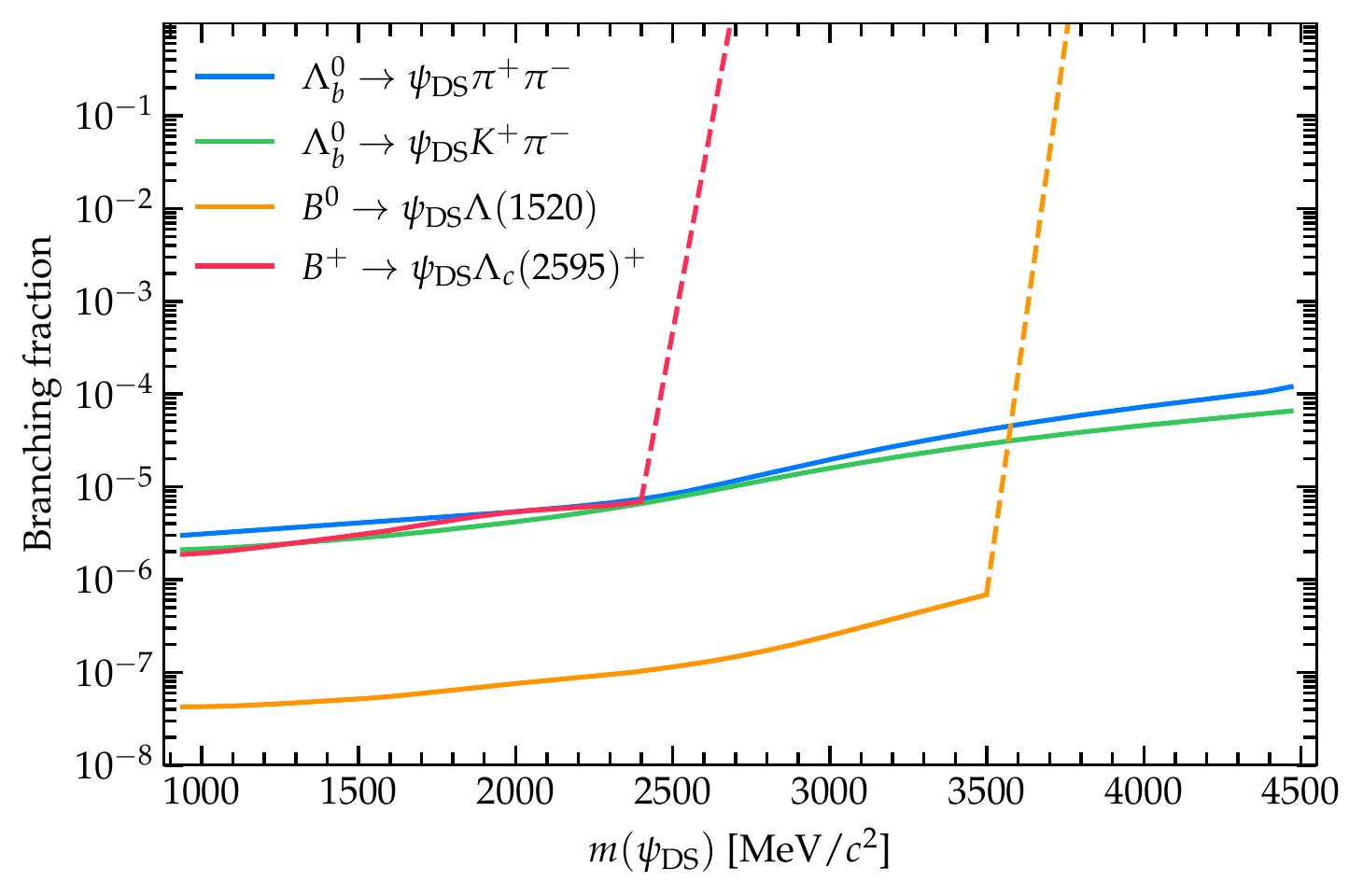}
\includegraphics[width=0.49\textwidth]{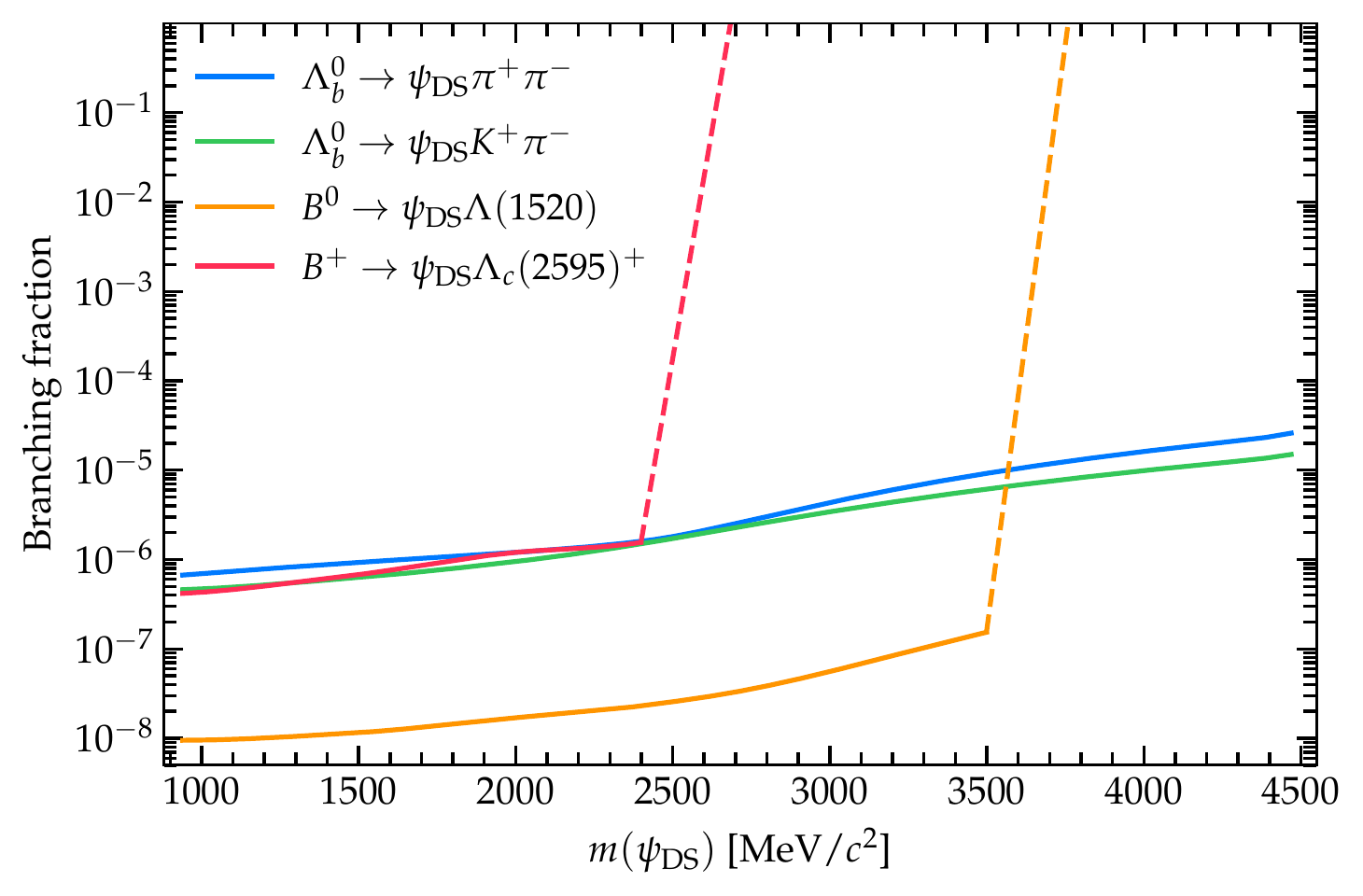}
\caption{Projected upper limits at $95\,\%$ CL that are expected to be reached with the LHCb experiment with a data set corresponding to integrated luminosities of (left) $15\,\text{fb}^{-1}$ and (right) $300\,\text{fb}^{-1}$.}
\label{fig:limits_baryoDM}
\end{figure}

Recently a model was proposed that connects DM production with a $B$-mesogenesis mechanism to explain the baryon asymmetry in the universe~\cite{DarkBaryons1, DarkBaryons2}. A neutral $B$ meson can oscillate and then decay $C\!P$-violating into a DS baryon $\psi_{\text{DS}}$ and SM hadrons. The baryon asymmetry $Y_B$ can then be written as 
\begin{align}
Y_B\propto A^{s,d}_{\text{SL}}\times\mathcal{B}(b\to\psi_\text{DS}X),
\end{align}
where $A^{s,d}_{\text{SL}}$ denotes the semileptonic asymmetry and $X$ a hadronic final state that balances out the baryon number in the $b\to\psi_\text{DS}X$ decay.
In order for the mechanism to work, the semileptonic asymmetry has to be positive. Intense efforts are ongoing to measure the parameters $A^s_{\text{SL}}$ and $A^d_{\text{SL}}$ directly~\cite{HFLAV} and indirectly~\cite{UTFit} with main contributions from the LHCb experiment. To fulfill the observed baryon asymmetry, the inclusive branching fraction of $b\to\psi_\text{DS}X$ has to be larger than $10^{-4}$ with experimental upper limits between $\leq10^{-4}$ and $10^{-2}$ depending on the mass of the $\psi_\text{DS}$ particle~\cite{DarkBaryons2,ALEPH}. Exclusive decays cannot be lower than the $10^{-6}$ level. The Belle collaboration recently published a search for one of the possible decay modes, $B^0\to\Lambda$ decays plus missing energy, leading to limits below $\mathcal{O}(10^{-5})$~\cite{BelleBToLambda}.
\begin{sloppypar}
A feasibility study has been performed for the search for such baryon-number violating $b$-hadron decays with the LHCb experiment~\cite{DarkBaryonProspects}.
The decays \mbox{$B^0 \rightarrow \psi_{\text{DS}} \Lambda(1520) (\rightarrow pK^-)$},
\mbox{$B^+ \rightarrow \psi_{\text{DS}} \Lambda_c(2595)^+ (\rightarrow \pi^+ \pi^- \Lambda_c^+(\rightarrow p K^- \pi^+ ))$}, \mbox{$\Lambda_b^0 \rightarrow \psi_{\text{DS}} K^+\pi^-$}, and \mbox{$\Lambda_b^0 \rightarrow \psi_{\text{DS}}\pi^+\pi^-$} are investigated, which are sensitive to different underlying quark-level transitions.
Events, generated with \textsc{Pythia} 8~\cite{Pythia} and propagated through a simplified detector simulation~\cite{DetSim} are analysed with a combination of kinematic and particle identification requirements, the closeness of the SM final state particles, the isolation of the final states with respect to the remainder of the event, as well as the missing momentum transverse to the $b$-hadron direction of flight.
Under the assumption of known background, the expected sensitivities of these searches with benchmarks of integrated luminosities of $15\,\text{fb}^{-1}$ (expected to be collected in the coming data taking period until 2024) and $300\,\text{fb}^{-1}$ (expected to correspond to the full LHCb data set by the end of the LHC) are shown in Fig.~\ref{fig:limits_baryoDM}.
They demonstrate the ability of the LHCb experiment to explore the full parameter space of the proposed model.
However, achieving the full sensitivity depends on a precise knowledge of the background level with systematic uncertainties of below $0.1\,\%$.
One way to improve the background suppression and thus reduce the reliance on the background level systematics would be the reconstruction of the $b$-hadrons through $\Sigma^+_b\to\Lambda^0_b\pi^+$ or $B_{s2}^{\ast 0}\rightarrow B^+ K^-$ decays. This would allow the precise reconstruction of the $\psi_{\text{DS}}$ mass and provide further handles to discriminate the signal from background.
\end{sloppypar}

\section{Conclusions and outlook}
The LHCb experiment has a unique coverage to search for Dark Matter particles especially in the low mass and significant displacement region. This results in several world-best limits, e.g. in the Dark Photon and inclusive $X\to\mu^+\mu^-$ searches described in this document. One of the keys to these successes is the use of fully online-selected data, which allows a larger physics reach with smaller data sizes. Currently the LHCb experiment is being upgraded, starting to take data in 2022.
The upgraded data set provides bright prospects and new tools for Dark Matter searches.
Due to the large data size with integrated luminosities up to $300\,\text{fb}^{-1}$ at centre-of-mass energies of $14\,\text{TeV}$, the experiment will be able to probe the full parameter region of models like the baryonic Dark Matter proposal discussed in these proceedings. Furthermore, the use of a fully software-based trigger system will allow to explore many more signatures~\cite{Stealth}.


\begin{thebibliography}{99}
\bibitem{DM2016}
J. Alexander et al., \href{https://arxiv.org/abs/1608.08632}{[\tt arXiv:1608.08632 (hep-ph)]}
\bibitem{DarkPhoton1}
LHCb collaboration: R. Aaij et al., \href{https://doi.org/10.1103/PhysRevLett.120.061801}{PRL 120 (2018) 6, 061801} \href{https://arxiv.org/abs/1710.02867}{[\tt arXiv:1710.02867 (hep-ex)]}
\bibitem{DarkPhoton2}
LHCb collaboration: R. Aaij et al., \href{https://doi.org/10.1103/PhysRevLett.124.041801}{PRL 124 (2020) 4, 041801} \href{https://arxiv.org/abs/1910.06926}{[\tt arXiv:1910.06926 (hep-ex)]}
\bibitem{Trigger_Performance}
LHCb collaboration: R. Aaij et al., \href{https://doi.org/10.1088/1748-0221/14/04/P04013}{JINST 14 (2019) 04, P04013} \href{https://arxiv.org/abs/1812.10790}{[\tt arXiv:1812.10790 (hep-ex)]}
\bibitem{DarkPhoton_e}
P. Ilten et al., \href{https://doi.org/10.1103/PhysRevD.92.115017}{PRD 92, 115017 (2015)} \href{https://arxiv.org/abs/1509.06765}{[\tt arXiv:1509.06765 (hep-ph)]}
\bibitem{Xmumu}
LHCb collaboration: R. Aaij et al., \href{https://doi.org/10.1007/JHEP10(2020)156}{    JHEP 10 (2020) 156} \href{https://arxiv.org/abs/2007.03923}{[\tt arXiv:2007.03923 (hep-ex)]}
\bibitem{2HDM}
G.C. Branco et al., \href{https://doi.org/10.1016/j.physrep.2012.02.002}{PR 516 (2012) 1-102} \href{https://arxiv.org/abs/1106.0034}{[\tt arXiv:1106.0034 (hep-ex)]}
\bibitem{CMS_mmb}
CMS collaboration:     A.M. Sirunyan et al., \href{https://doi.org/10.1007/JHEP11(2018)161}{    JHEP 11 (2018) 161} \href{https://arxiv.org/abs/1808.01890}{[\tt arXiv:1808.01890 (hep-ex)]}
\bibitem{DarkBaryons1}
G. Elor et al., \href{https://doi.org/10.1103/PhysRevD.99.035031}{PRD 99 99 (2019) 3, 035031} \href{https://arxiv.org/abs/1810.00880}{[\tt arXiv:1810.00880 (hep-ph)]}
\bibitem{DarkBaryons2}
G. Alonso-Álvarez et al., \href{https://doi.org/10.1103/PhysRevD.104.035028}{PRD 104 (2021) 3, 035028} \href{https://arxiv.org/abs/2101.02706}{[\tt arXiv:2101.02706 (hep-ph)]}
\bibitem{HFLAV}
HFLAV collaboration: Y. Amhis et al., \href{https://doi.org/10.1140/epjc/s10052-020-8156-7}{EPJC (2021) 81} \href{https://arxiv.org/abs/1909.12524}{[\tt arXiv: 1909.12524 (hep-ex)]}, updated results and plots available at \href{https://HFLAV.web.cern.ch/}{{\texttt{https://HFLAV.web.cern.ch/}}}
\bibitem{UTFit}
UTfit collaboration: M. Bona et al., \href{https://doi.org/10.1103/PhysRevLett.97.151803}{PRL 97 (2006) 151803} \href{https://arxiv.org/abs/hep-ph/0605213}{[\tt arXiv: hep-ph/0605213 (hep-ph)]}, updated results and plots available at \href{http://www.utfit.org/UTfit/WebHome}{{\texttt{http://www.utfit.org/}}}
\bibitem{ALEPH}
ALEPH collaboration: R. Barate et al., \href{https://doi.org/10.1007/s100520100612}{EPJC 19 (2001) 213-227} \href{https://arxiv.org/abs/hep-ex/0010022}{[\tt arXiv:hep-ex/0010022 (hep-ex)]}
\bibitem{BelleBToLambda}
Belle collaboration: C. Hadjivasiliou et al., \href{https://arxiv.org/abs/2110.14086}{[\tt arXiv: 2110.14086 (hep-ex)]}
\bibitem{DarkBaryonProspects}
A. Brea Rodríguez et al., \href{https://arxiv.org/abs/2106.12870}{[\tt 2106.12870 (hep-ph)]}
\bibitem{Pythia}
T. Sj\"ostrand et al., \href{https://doi.org/10.1016/j.cpc.2015.01.024}{CPC 191 (2015) 159-177} \href{https://arxiv.org/abs/1410.3012}{[\tt arXiv: 1410.3012 (hep-ph)]}
\bibitem{DetSim}
V. Chobanova et al.,\href{https://arxiv.org/abs/2012.02692}{[\tt arXiv:2012.02692 (hep-ex)]}
\bibitem{Stealth}
M. Borsato et al., \href{https://arxiv.org/abs/2105.12668}{[\tt arXiv:2105.12668 (hep-ph)]}
\end{thebibliography}
\end{document}